\documentclass{article}
\usepackage{graphicx} 
\usepackage{amsmath}
\usepackage{amssymb}
\usepackage{booktabs}
\usepackage{geometry}
\usepackage[colorlinks=true, linkcolor=red, citecolor=blue, urlcolor=black]{hyperref}
\newtheorem{theorem}{Theorem}

\title{The virial theorem and the Price equation}
\author{Catherine Felce\thanks{Division of Physics, Mathematics and Astronomy, California Institute of Technology, Pasadena, CA}, Steinunn Liorsd\'{o}ttir\thanks{Flintridge Preparatory School, La Ca\~{n}ada, CA} \ and Lior Pachter\thanks{Division of Biology and Biological Engineering, and Department of Computing and Mathematical Sciences, California Institute of Technology, Pasadena, CA, \href{mailto:lpachter@caltech.edu}{lpachter@caltech.edu}}}
\date{\today}

\begin{document}

\maketitle

\begin{abstract}
We observe that the time-averaged continuous Price equation is identical to the positive momentum virial theorem, and we discuss the applications and implications of this connection.
\end{abstract}

\section*{The virial theorem}
The virial theorem was first described by Rudolf Clausius in connection with his studies on heat transfer \cite{clausius1870xvi}. In its simplest form, it relates the time-averaged kinetic energy $\langle T \rangle_{\tau} = \langle \frac{1}{2}\sum_{i=1}^n m_i v_i(t)^2 \rangle_{\tau}=\frac{1}{\tau} \int_{0}^{\tau} \frac{1}{2}\sum_{i=1}^n m_i v_i(t)^2 dt$ of $n$ objects with masses $m_1,\ldots,m_n$ and velocities $v_1(t),\ldots,v_n(t)$, to their time averaged potential energy $\langle U \rangle_{\tau} = \langle \sum_{i=1}^n F_i(t) z_i(t)\rangle_{\tau}$, where $F_1(t),\ldots,F_n(t)$ are the forces acting on the $n$ objects and $z_1(t),\ldots,z_n(t)$ are their respective positions:
\begin{theorem}[Virial theorem, 1870]
For stably bound gravitational systems, 
\begin{equation}
\label{eq:virialtheorem}
\langle T \rangle_{\tau} =  -\frac{1}{2} \langle U\rangle_{\tau}.
\end{equation}
\end{theorem} 
The mathematical underpinning of the virial theorem is the product rule from calculus. The derivative of the Clausius virial $S(t) = \sum_{i=1}^n p_i(t)z_i(t)$ where $p_i(t) = m_iv_i(t)$ is
\begin{eqnarray}
 \frac{dS(t)}{dt} & = & \sum_{i=1}^n m_i \frac{dv_i(t)}{dt} z_i(t) + \sum_{i=1}^n p_i(t) \frac{dz_i(t)}{dt} \nonumber \\
& = &  \sum_{i=1}^n F_i(t)z_i(t) + \sum_{i=1}^n m_iv_i(t)^2 \nonumber \\
& = & U + 2T,  \nonumber \\
\implies \left\langle \frac{dS(t)}{dt} \right\rangle_{\tau} & = &  \langle U \rangle_{\tau} + 2\langle T \rangle_{\tau}. \label{eq:virialfull}
\end{eqnarray}
 Since for stably bound systems the velocities and positions of objects have upper and lower bounds, the average of the derivative of $S(t)$ over a period of time $\tau$ will be zero in the limit of large $\tau$, i.e. for large $\tau$, $\langle \frac{dS(t)}{dt} \rangle_{\tau}  \approx 0$. When $\left\langle \frac{dS(t)}{dt} \right\rangle _{\tau}= 0$, we obtain from (\ref{eq:virialfull}) the virial theorem (\ref{eq:virialtheorem}): $\langle T \rangle_{\tau} =  -\frac{1}{2} \langle U\rangle_{\tau}$.

The virial theorem was well known to physicists in the late 19th and early 20th centuries \cite{rayleigh1905xlii,einstein1922grundlage}, however, its power as a discovery tool for astrophysics was first highlighted by Fritz Zwicky \cite{zwicky1933rotverschiebung}. Zwicky used the virial theorem to estimate the mass of the Coma cluster, thereby identifying a mass deficit in comparison to luminosity estimates, leading him to posit the existence of what he called {\em dunkle materie} (dark matter) \cite{zwicky1933rotverschiebung}. Although Zwicky's mass estimates were inaccurate \cite{the1986aj,merritt1987distribution}, the principle of using the virial theorem to identify a measurement gap was sound, and the virial theorem has become widely used in physics and astrophysics. It can be used to derive classic laws such as the ideal gas law \cite{fowler1967statistical,hanson1995virial}, and extensions are applicable in many settings, including quantum mechanics \cite{georgescu1999virial}, astrophysical hydrodynamics \cite{shore2012introduction}, and fluid mechanics \cite{oguz1990generalization, alazard2023virial}.

\section*{The Price equation}
The Price equation \cite{price1970selection} pertains to selection in evolutionary processes. It was motivated by a desire to understand the evolution of altruism \cite{harman2011price}, and has been described as a ``fundamental theorem of evolution" \cite{queller2017fundamental} due to its generalization and unification of many results in evolutionary biology. For example, Fisher's fundamental theorem of natural selection \cite{fisher1930genetical} is a special case of the Price equation \cite{queller2017fundamental,frank1997price}. 

The Price equation relates the change in a trait in a population over time, to fitness values in subpopulations. Formally, the (discrete) Price equation as published in \cite{price1970selection} (we follow notation from \cite{frank1997price}) considers a numerical trait in $n$ subpopulations at time $t$ denoted ${\bf z}(t) = (z_1(t),\ldots,z_n(t))$. The subpopulations have sizes $p_1(t),\ldots,p_n(t)$, and have (Wrightian) fitness ${\bf w}(t) = (w_1(t),\ldots,w_n(t))$ defined by $w_i(t) = \frac{p_i(t+\Delta t)}{p_i(t)}$ where $\Delta t$ denotes the time interval of one generation \cite{wagner2010measurement}. Let $q_i(t) = \frac{p_i(t)}{\sum_{j=1}^n p_j(t)}$ be the relative size of the $i^{\rm th}$ population, and define the population average fitness to be $\overline{{\bf w}}(t) = \sum_{i=1}^n q_i(t)w_i(t)$. Note that ${\bf q}(t)$ forms a probability distribution for ${\bf w}(t)$ viewed as a random variable, and $\mathbb{E}({\bf w}(t)) = \overline{{\bf w}}(t)$. Let $\Delta z_i(t) = z_i(t+\Delta t) - z_i(t)$, $\Delta {\bf z}(t) = {\bf z}(t + \Delta t) - {\bf z}(t)$, and $\overline{{\bf z}}(t) = \sum_{i=1}^nq_i(t)z_i(t)$ with $\Delta \overline{{\bf z}}(t) = \overline{{\bf z}}(t+\Delta t) - \overline{{\bf z}}(t)$.

\begin{theorem}[The Price equation, 1970]
\begin{equation}
     \Delta \overline{{\bf z}}(t) = \frac{1}{\overline{{\bf w}}(t)}\mathrm{cov}({\bf w}(t),{\bf z}(t)) + \frac{1}{\overline{{\bf w}}(t)}\mathbb{E}({\bf w}(t) \odot \Delta {\bf z}(t)), \label{eq:discreteprice}
\end{equation} 
     where $\mathbb{E}({\bf w}(t) \odot \Delta {\bf z}(t))$ is the expected value of the Hadamard product of ${\bf w}(t)$ and $\Delta {\bf z}(t)$ with respect to the relative subpopulation sizes, and  $\mathrm{cov}({\bf w}(t),{\bf z}(t)) = \mathbb{E}({\bf w} \odot {\bf z}) - \mathbb{E}({\bf w}) \mathbb{E}({\bf z})$ is the covariance between the subpopulation fitnesses and trait values with respect to the relative subpopulation sizes.
     \end{theorem}

Intuitively, if subpopulation fitness has positive covariance with trait values, then the trait is beneficial, and the trait value, averaged across populations, will increase after a generation. However, if the covariance between subpopulation fitness and trait values is negative, higher trait values are detrimental and the trait value averaged across populations will decrease after a generation.

The Price equation as published in \cite{price1970selection} is discrete in time, and proof of the identity uses basic properties of expectation and covariance along with the fact that $q_i(t+\Delta t) = \frac{q_i(t)w_i(t)}{\overline{{\bf w}}(t)}$, which we leave as an exercise for the reader. Note that 
\begin{eqnarray*}
 \overline{{\bf w}}(t) \Delta \overline{{\bf z}}(t) & = & \overline{{\bf w}}(t) \overline{{\bf z}}(t+\Delta t)-\overline{{\bf w}}(t)\overline{{\bf z}}(t)\\
 & = & \overline{{\bf w}}(t) \sum_{i=1}^n q_i(t+\Delta t)z_i(t+\Delta t) - \overline{{\bf w}}(t)\overline{{\bf z}}(t)\\
 & = & \sum_{i=1}^n q_i(t) w_i(t) z_i(t+\Delta t) -\overline{{\bf w}}(t)\overline{{\bf z}}(t)\\
 & = &  \sum_{i=1}^n q_i(t)w_i(t)z_i(t) - \overline{{\bf w}}(t)\overline{{\bf z}}(t) + \sum_{i=1}^n q_i(t) w_i(t) z_i(t+\Delta t) - \sum_{i=1}^n q_i(t)w_i(t)z_i(t)\\
 & = & \mathrm{cov}({\bf w}(t),{\bf z}(t)) + \mathbb{E} ({\bf w}(t) \odot \Delta {\bf z}(t)),\\
 \implies \Delta \overline{{\bf z}}(t) & = & \frac{1}{\overline{{\bf w}}(t)}\mathrm{cov}({\bf w}(t),{\bf z}(t)) + \frac{1}{\overline{{\bf w}}(t)}\mathbb{E}({\bf w}(t) \odot \Delta {\bf z}(t)).
\end{eqnarray*}

The discrete-time Price equation has a continuous-time analog \cite{price1972fisher,ellner2011does}. It is formulated using the Malthusian fitness ${\bf r}(t) = r_1,\ldots,r_n$ given by $r_i(t) = \frac{1}{p_i(t)}\frac{dp_i(t)}{dt}$ instead of the Wrightian fitness ${\bf w}(t)$.
\begin{theorem}[The continuous Price equation, 1972]
\begin{equation}
    \frac{d}{dt} \mathbb{E}({\bf z}(t)) = \mathrm{cov}({\bf r}(t),{\bf z}(t))+ \mathbb{E}\left(\frac{d{\bf z}(t)}{dt}\right). \label{eq:contprice}
\end{equation}
\end{theorem}

The continuous-time Price equation (\ref{eq:contprice}) is the continuum limit of the discrete Price equation (\ref{eq:discreteprice}). To see this, we begin by multiplying the Price equation by $\frac{\overline{{\bf w}}(t)}{\Delta t}$:
\begin{eqnarray*}
     \frac{\overline{{\bf w}}(t)\Delta \overline{{\bf z}}(t)}{\Delta t} & = & \frac{1}{\Delta t}\mathrm{cov}({\bf w}(t),{\bf z}(t)) + \frac{1}{\Delta t}\mathbb{E}({\bf w}(t) \odot \Delta {\bf z}(t)).
\end{eqnarray*}
We will now see why, contrary to convention, we have indexed the variables in equation (\ref{eq:discreteprice}) with time. Starting with the left hand side, we observe that
\begin{eqnarray*}
    \lim_{\Delta t \rightarrow 0} \overline{{\bf w}}(t) & =  & \lim_{\Delta t \rightarrow 0} \sum_{i=1}^n q_i(t) w_i(t)\\
    & = & \lim_{\Delta t \rightarrow 0} \sum_{i=1}^n \frac{p_i(t)p_i(t+\Delta t)}{\left(\sum_{j=1}^np_j(t)\right)p_i(t)}\\
    & = & \frac{1}{\sum_{j=1}^np_j(t)} \lim_{\Delta t \rightarrow 0} \sum_{i=1}^n p_i(t+\Delta t)\, = \, 1.
    \end{eqnarray*}
Therefore, 
\begin{eqnarray*}
          \lim_{\Delta t \rightarrow 0} \frac{\overline{{\bf w}}(t)\Delta \overline{{\bf z}}(t)}{\Delta t} & = &  \lim_{\Delta t \rightarrow 0} \frac{\sum_{i=1}^n q_i(t+\Delta t)z_i(t+\Delta t) - \sum_{i=1}^nq_i(t)z_i(t)}{\Delta t}\\
          & = & \frac{d}{dt}\mathbb{E}({\bf z}(t)).
\end{eqnarray*}
The covariance term, in the limit as $\Delta t \rightarrow 0$, is given by
\begin{eqnarray*}
    \lim_{\Delta t \rightarrow 0} \frac{1}{\Delta t}\mathrm{cov}({\bf w}(t),{\bf z}(t)) & = &  \lim_{\Delta t \rightarrow 0}  \frac{\sum_{i=1}^n q_i(t)w_i(t)z_i(t)-\overline{{\bf w}}(t)\overline{{\bf z}}(t)}{\Delta t}.
\end{eqnarray*}
Let $g_i(t) = \frac{1}{\Delta t}ln(w_i(t))$. Note that $w_i(t)=e^{g_i(t)\Delta t}$ and that $\lim_{\Delta t \rightarrow 0}g_i(t) = r_i(t)$. Substituting $e^{g_i(t)\Delta t}$ for $w_i(t)$ yields
{\begin{eqnarray*}
    \lim_{\Delta t \rightarrow 0} \frac{1}{\Delta t}\mathrm{cov}({\bf w}(t),{\bf z}(t)) & = &  \lim_{\Delta t \rightarrow 0} \frac{\sum_{i=1}^n q_i(t) e^{g_i(t) \Delta t}z_i(t)-\sum_{i=1}^nq_i(t)e^{g_i(t)\Delta t}\overline{{\bf z}}(t)}{\Delta t}\\
    & = & \lim_{\Delta t \rightarrow 0} \frac{\sum_{i=1}^n q_i(t)e^{g_i(t)\Delta t}(z_i(t)-\overline{{\bf z}}(t))}{\Delta t}\\
    & = & \lim_{\Delta t \rightarrow 0} \sum_{i=1}^n q_i(t)g_i(t)e^{g_i(t)\Delta t}(z_i(t)-\overline{{\bf z}}(t)) \quad \mbox{by the L'H\^{o}pital-Bernoulli rule \cite{lhospitale1696}}\\
    & = &  \sum_{i=1}^n q_i(t)r_i(t)(z_i(t)-\overline{{\bf z}}(t))\\
    & = & \sum_{i=1}^n q_i(t)r_i(t) z_i(t) - \sum_{i=1}^nq_i(t) r_i(t)\overline{{\bf z}}(t)\\
    & = & \mathrm{cov}({\bf r}(t),{\bf z}(t)).
\end{eqnarray*}}
Finally, we have that
\begin{eqnarray*}
    \lim_{\Delta t \rightarrow 0} \frac{1}{\Delta t}\mathbb{E}({\bf w}(t) \odot \Delta {\bf z}(t)) & = & \lim_{\Delta t \rightarrow 0} \frac{ \sum_{i=1}^n q_i(t) e^{g_i(t)\Delta t} \Delta z_i(t)}{\Delta t} \\
    & = &  \sum_{i=1}^n q_i(t) \frac{dz_i(t)}{dt}\\
    & = & \mathbb{E}\left(\frac{d{\bf z}(t)}{dt}\right).
\end{eqnarray*}
In summmary, 
\begin{align*}
    \frac{\overline{{\bf w}}(t)\Delta {\bf z}(t)}{\Delta t} & \quad = \quad \frac{1}{\Delta t}\mathrm{cov}({\bf w}(t),{\bf z}(t)) + \frac{1}{\Delta t}\mathbb{E}({\bf w}(t) \odot \Delta {\bf z}(t)) & ( \mbox{discrete Price equation (\ref{eq:discreteprice}}))\\
    \Big\downarrow \scriptstyle{\lim_{\Delta t \rightarrow 0}} & \qquad\qquad\qquad\qquad \Big\downarrow \scriptstyle{\lim_{\Delta t \rightarrow 0}}   & \\
    \frac{d}{dt} \mathbb{E}({\bf z}(t)) & \quad = \quad \mathrm{cov}({\bf r}(t),{\bf z}(t)) + \mathbb{E}\left(\frac{d{\bf z}(t)}{dt}\right) & (\mbox{continuous Price equation (\ref{eq:contprice}}))
\end{align*}

\section*{The Price equation from the virial theorem}

In the physics setting, recall that the momentum $p_i(t)=m_iv_i(t)=m_i\frac{dz_i(t)}{dt}$. Let $r_i = \frac{1}{{p_i(t)}}\frac{dp_i(t)}{dt}$, i.e. acceleration divided by velocity. If all the momenta are greater than zero, i.e., $p_i(t)>0$ for all $i$, we can define relative momentum as $q_i(t)=\frac{p_i(t)}{\sum_{j=1}^np_j(t)}$. Consider the virial density $\tilde{S}(t) = \sum_{i=1}^n q_i(t)z_i(t)$ \cite{englert2014lectures}, whose derivative is $\frac{d\tilde{S}(t)}{dt} = \frac{d}{dt} \mathbb{E}({\bf z}(t))$. The product rule applied to the virial density is
\begin{eqnarray}
    \frac{d}{dt} \mathbb{E}({\bf z}(t)) & = & \sum_{i=1}^n\frac{d}{dt}\left(\frac{p_i(t)}{\sum_{j=1}^np_j(t)}\right) z_i(t) + \sum_{i=1}^n q_i(t) \frac{dz_i(t)}{dt} \nonumber \\
    & = & \sum_{i=1}^n \frac{\frac{dp_i(t)}{dt}\sum_{j=1}^np_j(t)- p_i(t)\sum_{j=1}^n\frac{dp_j(t)}{dt}}{(\sum_{j=1}^np_j(t))^2}z_i(t) + \mathbb{E}\left( \frac{d{\bf z}(t)}{dt} \right) \nonumber \\
     & = & \sum_{i=1}^n  \frac{p_i(t)}{\sum_{j=1}^np_j(t)}\left( r_i(t) - \frac{\sum_{j=1}^nr_j(t)p_j(t)}{\sum_{j=1}^np_j(t)} \right)z_i(t) + \mathbb{E}\left( \frac{d{\bf z}(t)}{dt} \right) \nonumber \\ 
     & = & \sum_{i=1}^n  q_i(t)r_i(t)z_i(t) - \sum_{j=1}^n q_j(t)r_j(t)\sum_{i=1}^n q_i(t) z_i(t) + \mathbb{E}\left( \frac{d{\bf z}(t)}{dt} \right) \nonumber \\ 
      & = & \mathbb{E}({\bf r}(t) \odot {\bf z}(t)) - \mathbb{E}({\bf r}(t))\mathbb{E}({\bf z}(t))+ \mathbb{E}\left( \frac{d{\bf z}(t)}{dt} \right), \nonumber \\ 
   \implies \quad \frac{d}{dt} \mathbb{E}({\bf z}(t)) & = & \mathrm{cov}({\bf r}(t),{\bf z}(t)) + \mathbb{E}\left( \frac{d{\bf z}(t)}{dt} \right). \label{eq:virialprice}
\end{eqnarray}

This shows that the virial (density) equation (\ref{eq:virialprice}) and the continuous Price equation (\ref{eq:contprice}) are mathematically identical. Therefore, the relationships between traits and fitness in evolutionary biology are not only reminiscent of the relationships between physical quantities like distance, velocity, and acceleration; they are the same. It is therefore not surprising to find a direct analog of the virial theorem in genetics \cite{frank1990distribution}: 
\begin{theorem}[Frank and Slatkin, 1990]
For a population in equilibrium where $i$ is the allelic state of a haploid genotype, $z_i=ci$ for some $c$, and ${\bf z}(t)^2 = {\bf z}(t) \odot {\bf z}(t)$,
\begin{equation*}
\mathrm{cov}({\bf w}(t),{\bf z}(t)^2) = -\mathbb{E}({\bf w} \odot \Delta  {\bf z}(t)^2). \label{eq:frank_slatkin}
\end{equation*}
\end{theorem}
In other words, the rate at which selection removes phenotypic variance from the population (left-hand side) is equal to the rate at which mutation adds variance (right-hand side). Other versions of the Price equation can provide physical insight.


In summary, we have the following relationships: 
\begin{align*}
\mbox{{\bf Biology}} &  \quad \qquad \qquad  \mbox{{\bf Physics}}\\
    \mbox{Price equation} & \quad \qquad \qquad \mbox{virial theorem} \\
    \Big\downarrow \scriptstyle{\lim_{\Delta t \rightarrow 0}} & \qquad\qquad\qquad \Big\downarrow \scriptstyle{p_i > 0}   \\
    \mbox{continuous Price equation} &  \qquad = \qquad  \mbox{positive momentum virial theorem}\\
\end{align*}
\vskip -0.15in
In genetics, the natural time increment to consider is discrete (generation), whereas in physics continuous-time is more natural. Thus, the discrete Price equation pertains to change in a trait after a single generation, whereas the virial theorem is formulated with continuous-time, and is additionally time averaged. However, the less intuitive forms of these equations that arise from the correspondences derived above may yield important insights. For example, the perspective of the virial theorem as a special case of the equipartition theorem \cite{podio2019virial} may be fruitful in evolutionary biology \cite{nourmohammad2013universality}. Translation between biology and physics via the virial theorem and the Price equation may also accelerate discovery of generalizations. While the stochastic Price equation in evolution \cite{rice2008stochastic} and the stochastic virial theorem in astronomy \cite{cresson2021stochastic} were discovered independently, their similarity suggests other generalizations could similarly parallel each other. Moreover, the virial theorem has been applied in a variety of fields (for example economics \cite{andersen2004population}), meaning that understanding its relationship to the Price equation could be relevant beyond physics and biology. 

\section*{Simple harmonic motion}

In the book ``Ecological Orbits: How Planets Move and Populations Grow" \cite{Ginzburg_Colyvan_2004}, Ginzburg and Colyvan show that a maternal effect can generate population cycles \cite{Ginzburg_94}, obviating the need for predator-prey models to explain such dynamics.
Formally, they posit that a trait $z(t)$ that changes over time is linked to the size $p(t)$ of a population as follows:
\begin{eqnarray}
    p({t+1)} & = & p(t) f(z(t)) \label{eq:fit} \label{eq:Ginzburg1} \\
    z(t+1) &  = & z(t) g \bigg(\frac{R}{p(t+1)}\bigg) 
    \textnormal{,} \label{eq:Ginzburg2}
\end{eqnarray}
where $f$ and $g$ are monotonically increasing functions (note that we have adjusted their notation). The maternal effect is captured by $z(t)$ on the right-hand side of the trait evolution equation, indicating that a trait associated with individuals in a generation depends on the trait of mothers in the current generation, as well as the fraction of the total resources $R$ available to each individual in the next generation, i.e. $\frac{R}{p(t+1)}$. We extend this model to include an additional function $h$ that captures the potentially dominant impact of transgenerational effects as seen in matrotrophic species \cite{Bian_2015,Harding_2001, reznick1995maternal,Roseboom2006DutchFamine}:

\begin{align}
    z(t+1) = z(t) g \bigg(\frac{R}{p(t+1)}\bigg) h \bigg(\frac{R}{p(t)}\bigg) 
    \textnormal{.}
\label{eq:new-mat}
\end{align}
This formulation captures environmental impacts on the mother which affect her offspring directly, such as nutrition during gestation \cite{thorne1976nutrition}. To derive a continuum limit from (\ref{eq:new-mat}) we replace $t+1$ with an arbitrary time increment $t+\Delta t$ to obtain
\begin{eqnarray}
    z(t+\Delta t) = z(t) \left(1+ \left( g \bigg(\frac{R}{p(t+\Delta t)}\bigg) h \bigg(\frac{R}{p(t)} \bigg) -1 \right) \Delta t\right)
    \textnormal{.}
\label{eq:new-mat-deltat}
\end{eqnarray}
Note that (\ref{eq:new-mat-deltat}) reduces to (\ref{eq:new-mat}) when $\Delta t = 1$. We can simplify (\ref{eq:new-mat-deltat}) in the case where transgenerational effects dominate, i.e., we assume that $g = 1$ and that 
\begin{eqnarray*}
    z(t+\Delta t) = z(t) \left(1+ \left( h \bigg(\frac{R}{p(t)} \bigg) -1 \right)\Delta t\right)
    \textnormal{.}
\label{eq:new-mat-simplified}
\end{eqnarray*}
In this case
\[
\frac{dz(t)}{dt} = \lim_{\Delta t \rightarrow 0} \frac{z(t+\Delta t)-z(t)}{\Delta t} = z(t) \left( h \bigg(\frac{R}{p(t)} \bigg) - 1 \right),
\]
or equivalently 
\begin{equation}
\frac{d}{dt}(\ln z(t)) =  h \bigg(\frac{R}{p(t)} \bigg)-1. \label{eq:dzdt}
\end{equation}
We note that it makes biological sense that $h$ should be a monotonically increasing concave function, since it captures the diminishing returns of increasing food per individual mother. This motivates the following specific functional form for $h$:
\begin{equation}
    h \bigg(\frac{R}{p(t)}\bigg) = \frac{1}{m} \ln{\bigg( \frac{R}{p(t)} \bigg)}, \label{eq:hform}
\end{equation}
where $\frac{1}{m}$ is a scaling factor representing the strength of the gestational maternal effect. Note that in what follows $h$ can be more general and include an additive constant, although we omit it for simplicity of presentation. Substituting (\ref{eq:hform}) into (\ref{eq:dzdt}) and defining $p_0 := \frac{R}{e^m}$ yields

\begin{equation}
    \frac{d\ln{z(t)}}{dt} = - \frac{1}{m} \left( \ln{p(t)} - \ln{p_0} \right).
\label{eq:mass}
\end{equation}


\begin{figure}[ht]
    \centering
    \includegraphics[width=0.8\textwidth]{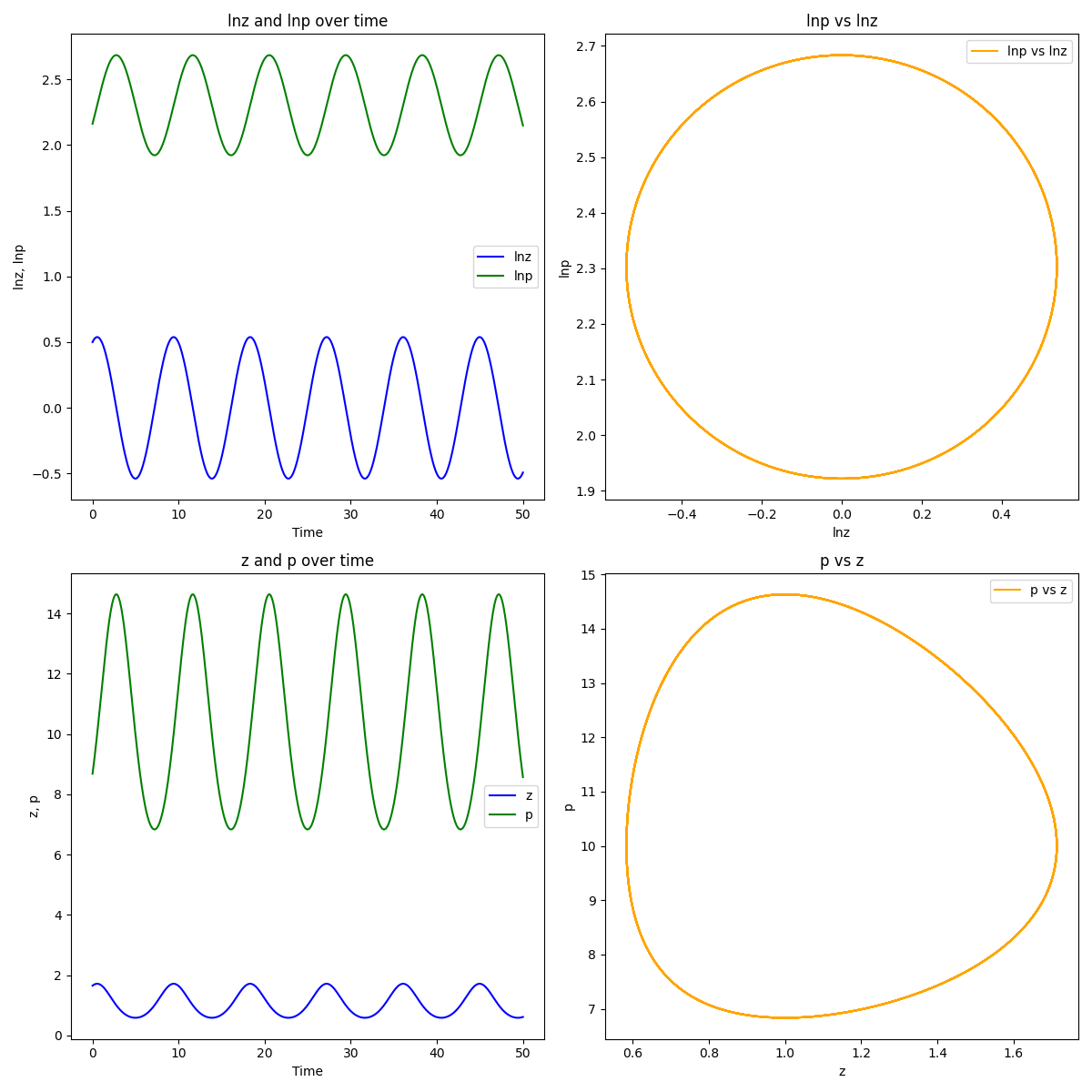}
    \caption{Simple harmonic motion of $\ln{z}$ and $\ln{p}$.}
    \label{fig:lnz_lnp}
\end{figure}

Now, as suggested by the first term in (\ref{eq:virialprice}), we consider the relationship between Malthusian fitness $r(t) = \frac{d \ln{p(t)}}{dt}$ and trait value $z(t)$. Similarly to (\ref{eq:dzdt}), the continuum limit of (\ref{eq:fit}) becomes
\begin{equation*}
    r(t) = \frac{d \ln{p(t)}}{dt} = f(z(t)) - 1.
\end{equation*}
If we further assume that fitness is linearly related to the logarithm of the trait value, i.e. $f(z(t))= k \textnormal{ln}(z(t)) + c$, for some constants, $c$ and $k$, we have that:
\begin{align}
\label{eq:log-fitness}
    \frac{d \textnormal{ln}(p(t))}{dt} = k \textnormal{ln}(z(t)) + c -1
    \textnormal{.}
\end{align}
Then, differentiating  (\ref{eq:mass}) and setting $z_0 := e^{\frac{1-c}{k}}$ we find that
\begin{equation*}
    \frac{d^2 \textnormal{ln}(z(t))}{dt^2} = -\frac{k}{m} (\textnormal{ln}(z(t)) - \textnormal{ln}(z_0)) 
    \textnormal{.}
\end{equation*}

Note that this is the second order differential equation describing simple harmonic motion (SHM) for the logarithm of the trait, where $k$ is the analog of a spring constant. The ``stiffness'' of the spring, $k$, is related to the strength of the trait's effect on fitness. The angular frequency of the motion is determined by the product of $k$ with the strength of the maternal effect, $\frac{1}{m}$, via $\omega=\sqrt{\frac{k}{m}}$.  In other words, \begin{equation*}
\ln z(t) = A \cos(\omega t) + B \sin(\omega t) + \ln z_0,
\end{equation*} 
where $A = \ln{z(0)} - \ln z_0$ and $B = \frac{1}{\omega}\left( \frac{d \ln{z(t)}}{dt}(0) \right)$.

The logarithmic population size also oscillates with SHM according to:
\begin{equation*}
    \frac{d^2\textnormal{ln}(p(t))}{dt^2} = - \frac{k}{m} (\textnormal{ln}(p(t)) - \textnormal{ln}(p_0)),
\end{equation*}
so that
\begin{equation*}
\ln p(t) = \left(-\frac{k B}{\omega}\right) \cos(\omega t) + \left(\frac{k A}{\omega}\right) \sin(\omega t) + \ln p_0.
\end{equation*}

Equation (\ref{eq:mass}) shows that, for a certain form of maternal effect, there is a natural relationship between the time derivative of (the logarithm of) the trait value and the logarithm of the population size.
By choosing the form of the selection ``force" via $f(z(t))$, we can consider different kinds of ``bound motion" of which simple harmonic motion is a fundamental example. 

\begin{figure}[ht]
    \centering
    \includegraphics[width=0.8\textwidth]{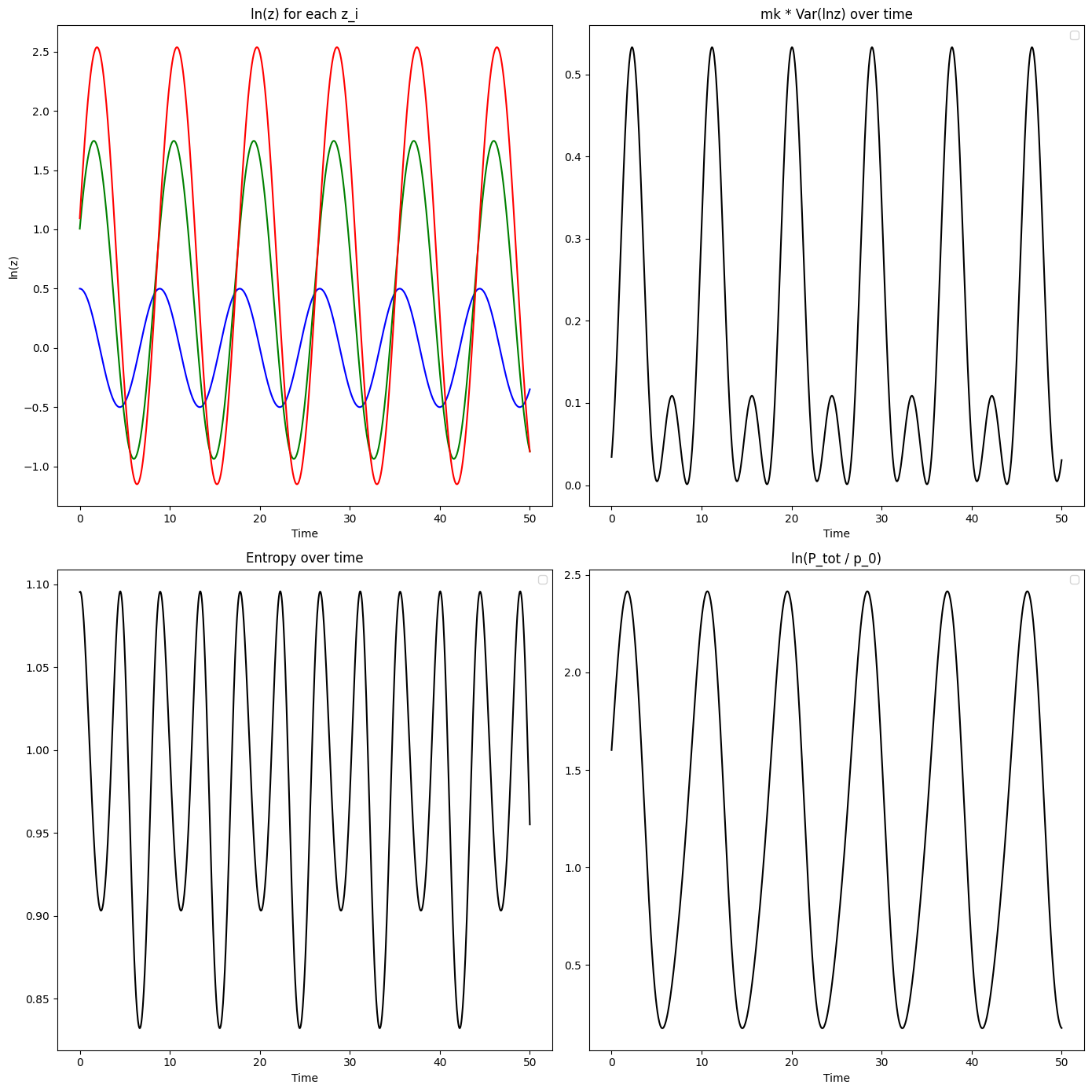}
    \caption{Behavior of three subpopulations and the quantities in (\ref{eq:virialsimple}).}
    \label{fig:fig2}
\end{figure}
The evolution equations (\ref{eq:Ginzburg1},\ref{eq:Ginzburg2}) and the extension (\ref{eq:new-mat}) deal with single populations but can be readily extended to multiple subpopulations, which can then be aggregated to shed light on the behavior of a full system. Consider the case in which the $i^{\rm th}$ subpopulation from among the $n$ subpopulations has a value for a trait represented by $z_i(t)$, a population size $p_i(t)$, a fitness scaling $k_i$, and individual maternal effects $m_i$. Suppose, in addition, that each subpopulation follows SHM as described above. The virial equation (\ref{eq:virialprice}) applied to $\ln{{\bf z}(t)}$ is

\begin{equation*}
    \frac{d}{dt} \mathbb{E} (\ln{{\bf z}(t)}) = \textnormal{cov}({\bf r}(t), \ln {\bf z}(t) + \mathbb{E} \left( \frac{d \ln{\bf z}(t)}{dt} \right)
    \textnormal{.}
\end{equation*}

 Since each subpopulation performs SHM in $\ln{z(t)}$, $\mathbb{E}[\textnormal{ln}(z)] = \sum_{i=1}^n q_i \ln{z_i}$ is bounded, so the time average of $\frac{d}{dt} \mathbb{E} (\ln{{\bf z}(t)})$ will go to zero. From equations (\ref{eq:log-fitness}) and (\ref{eq:mass}), we therefore have:

\begin{eqnarray}
    0 & = & \left\langle \textnormal{cov}({\bf k} \odot \ln{{\bf z}(t)} + {\bf c}, \ln{{\bf z}(t)} \right\rangle_{\tau} + \left\langle - \sum_i q_i(t) \frac{1}{m_i} (\textnormal{ln} (p_i(t)) -  \textnormal{ln} (p_{0i}(t))) \right\rangle_{\tau}\nonumber \\
    & = & \left\langle \textnormal{cov}({\bf k} \odot \ln{{\bf z}(t)} + {\bf c}, \ln{{\bf z}(t)} \right\rangle_{\tau} + \left\langle  -\sum_{i=1}^n \frac{1}{m_i}q_i(t) \ln{(q_i(t))} \right\rangle_{\tau}   -   \left\langle \sum_{i=1}^n \frac{1}{m_i} q_i(t) \ln{\left( \frac{P_{\textnormal{tot}}(t)}{p_{0i}}\right)} \right\rangle_{\tau}. \label{eq:virial_SHM}
\end{eqnarray} 

Note that the standard form of the virial theorem for SHM is an energy equation in $\langle -2U \rangle$ and $\langle 2T \rangle$, yielding the relation $\langle T \rangle= \langle U \rangle$. On the other hand, (\ref{eq:virial_SHM}) is a velocity equation. In the special case where the values $k_i$, $m_i$, $c_i$ and $p_{0i}$ are the same between subpopulations and given by $k$, $m$, $c$ and $p_0$, respectively, the above simplifies to

\begin{eqnarray}
mk\left\langle \textnormal{var}(\ln{{\bf z}(t)})    \right\rangle_{\tau} +\left\langle -\sum_{i=1}^n q_i(t) \textnormal{ln}(q_i(t)) \right\rangle_{\tau}  & = &   \left\langle  \ln{\left(\frac{P_{\textnormal{tot}}(t)}{p_0} \right)}  \right\rangle_{\tau}.
\label{eq:virialsimple}
\end{eqnarray}

This equation describes the balance between variation in the trait between subpopulations and the entropy in the distribution of subpopulation sizes, when subpopulations are following SHM in the way we have described. When subpopulations have different trait values, selection acts to create a non-uniform distribution of populations sizes. This illustrates the essence of the general case (\ref{eq:virial_SHM}), where the Shannon entropy is replaced by a weighted entropy \cite{suhov2016basic}.

Notably, the consideration of overall population dynamics resulting from multiple subpopulations exhibiting SHM, leads immediately to a universality observation stemming from the Fourier theorem \cite{rudin1964principles}:

\begin{theorem}[Universality of ecological orbits]
    Maternal effects driving simple harmonic motion in subpopulations are sufficient to generate any periodic population dynamics that satisfy the Dirichlet conditions.  
\end{theorem}

\section*{Evolutionary theory and Newtonian mechanics}
The {\em dynamical interpretation} of evolutionary theory posits a correspondence between theories of evolution and Newtonian mechanics \cite{elliott1984nature, hitchcock2014evolutionary}. In this framework, notions such as selection or mutation in biology are associated to forces in physics \cite{elliott1984nature}. The identical form of equations (\ref{eq:contprice}) and (\ref{eq:virialprice}) can constrain such associations and clarify subsequent analogies. For example, although the standard form of the virial theorem in (\ref{eq:virialfull}) is an energy equation, equation (\ref{eq:virialprice}) is a velocity equation, which in biology translates to rates of change of biological quantities. Furthermore, rate of change of a trait or phenotype, i.e., $\frac{d}{dt} \mathbb{E}({\bf z}(t))$ in equation  (\ref{eq:contprice}) or the finite difference $\Delta {\bf \overline{z}}(t)$ in equation (\ref{eq:discreteprice}), corresponds to the momentum-averaged bulk velocity of a collection of physical objects. The remaining terms in each equation similarly have physical meaning in the context of classical mechanics in equation (\ref{eq:virialprice}), or biology in equation (\ref{eq:discreteprice}).

\begin{table}[ht]
\centering
\caption{Glossary of Terms}
\label{tab:glossary}
\begin{tabular}{llll}
\toprule
Variable & Biology & Physics & Physical Units \\
\midrule
$i$ & subpopulation & object & 1 \\
$ z_i(t) $ & trait / phenotype & position & $m$ \\
$ \frac{dz_i(t)}{dt} $ & evolutionary rate & velocity & $m \cdot s^{-1}$ \\
$ r_i(t) $ & Malthusian fitness & acceleration $\div$ velocity & $s^{-1}$ \\
$ p_i(t) $ & population size & momentum & $kg \cdot m \cdot s^{-1}$ \\
$ q_i(t) $ & relative population size & relative momentum & 1 \\
$ \frac{dp_i(t)}{dt} $ & population growth rate & force & $kg \cdot m \cdot s^{-2}$ \\
$ \mathbb{E}({\bf z}(t))$ & population-averaged trait / phenotype & momentum-averaged position & $m$ \\
$ \frac{d}{dt}\mathbb{E}({\bf z}(t)) $ & group evolutionary rate & bulk momentum velocity & $m \cdot s^{-1}$ \\
$\mathrm{cov}({\bf r}(t),{\bf z}(t))$ & selection rate & extrinsic momentum velocity & $m \cdot s^{-1}$ \\
$ \mathbb{E}\left(\frac{d{\bf z}(t)}{dt}\right) $ & transmission rate & intrinsic momentum velocity & $m \cdot s^{-1}$ \\
\bottomrule
\end{tabular}
\end{table}

The momentum-averaged positions $\mathbb{E}({\bf z}(t))$ and velocities $\frac{d}{dt}\mathbb{E}({\bf z}(t))$ are discrete analogs of momentum-averaged position and momentum velocity in electromagnetism, where they emerge from the virial density in an application of the virial theorem to electromagnetic pulses \cite{englert2014lectures}. Whereas $\mathrm{cov}({\bf r}(t),{\bf z}(t))$ is frequently referred to as the {\em selection term} in the Price equation \cite{bourrat2023price}, the connection to the virial theorem suggests that it is better described as a selection {\em rate} (Table \ref{tab:glossary}). Similarly, the {\em transmission} term $\mathbb{E}\left(\frac{d{\bf z}(t)}{dt}\right)$ is more accurately a transmission {\em rate}. Most significantly, while the dynamical interpretation typically relies on associating force to natural selection, drift, migration, or mutation \cite{hitchcock2014evolutionary}, the equivalence between the virial theorem and the Price equation, suggests that force is more naturally associated to fitness. The correspondence of force to a rate of change is not surprising, since force is also a rate of change, specifically the rate of change of momentum. This stands more in line with the {\em statistical interpretation} of evolutionary theory \cite{walsh2000chasing,walsh2002trials,matthen2002two}, which, among several critiques of the dynamical interpretation, finds fault with the analogies of biological processes such as mutation with forces in physics, arguing that the physical forces are causal in a way that processes such as selection or mutation are not \cite{walsh2000chasing}. However, the analogy of population growth with force can be viewed as consistent with the dynamical interpretation; for example, population growth can directly affect DNA polymorphism patterns \cite{williamson2005simultaneous}. Moreover, the virial theorem in the setting of the ecological simple harmonic oscillator affirms \cite{hitchcock2014evolutionary} in noting that ``natural selection turns out to be more similar to forces such as friction and elastic forces rather than the more canonical gravitation."

In particular, considerations of the analogies between biology and physics via the virial theorem led us to generalize the work of \cite{Ginzburg_Colyvan_2004} and to derive the ecological simple harmonic oscillator, which to our knowledge is the first example of such an oscillator that emerges solely from maternal effects and does not require a predator-prey or other more sophisticated model. The extension of (\ref{eq:Ginzburg2}) to (\ref{eq:new-mat}) is interesting in its own right, and should be interesting to develop in future work. Moreover, Theorem 5 shows that subpopulations subject to distinct maternal effects can generate arbitrarily complex population dynamics, thereby affirming the main thesis of \cite{Ginzburg_Colyvan_2004}.

Ultimately, analogies between the Price equation and the virial theorem point towards potentially productive directions for exploration in both biology and physics. The statistical framing of the virial theorem in (\ref{eq:virialprice}) highlights phenomena that may have been overlooked in the physics realm. For example, the first term on the right-hand side of (\ref{eq:virialprice}), namely $\mathrm{cov}({\bf r}(t),{\bf z}(t))$, can be understood to quantify the extent of the Yule-Simpson effect \cite{pearson1899vi,yule1903notes,simpson1951interpretation}, which describes a situation where within-group trends can be reversed upon averaging. In biology, the Price equation has the potential to be used more widely as a tool. Although it has been hailed as a unifying framework for researchers \cite{luque2017one}, one that ``can serve as a heuristic principle to formulate and systematize different theories and models in evolutionary biology" \cite{luque2021mirror}, the emphasis on its use has been more oriented toward understanding how it generalizes specific equations, rather than applying it for biological discovery. For example, the Price equation can be used to derive the Breeder's equation \cite{bijma2020price, zhang2010change}, Fisher's fundamental theorem \cite{queller2017fundamental,price1972fisher}, the house of cards approximation for genetic variance at mutation-selection balance \cite{zhang2010change,turelli1984heritable}, and many other formulas and identities in genetics \cite{zhang2010change, rice2004evolutionary}. However, it has been referred to as a tautology and a vacuous statement without application. In \cite{van2012group} the Price equation is described as a theorem that establishes that ``If the left-hand side is computed as suggested in \cite{price1970selection}, and the right-hand side too, then they are equal." This critique of the Price equation, namely that it does not and cannot serve as a {\em tool}, stands in contradiction to evidence from physics, where the mathematically equivalent virial theorem has been understood as a powerful tool since its use to discover dark matter in 1933 \cite{zwicky1933rotverschiebung}. The manifold applications of the virial theorem \cite{marc1985virial} suggest that there is still much to gain from the application of the Price equation as a tool for biology. In fact, the equivalence we have demonstrated between the Price equation and the virial theorem shows that the description of missing heritability as dark matter \cite{manolio2009finding} may be understood to be more than just an informal analogy between mysteries in genetics and astronomy.


\section*{Author contributions}

SL studied the virial theorem while participating in the ``Introduction to Astrophysics" cluster in the COSMOS summer program held at UC Irvine from July 9, 2023 to August 4, 2023. Specifically, she used the virial theorem to repeat Zwicky's Coma cluster mass estimates using modern measurements of velocity dispersion and galaxy positions. LP learned of the virial theorem from SL and, in discussing its proof with SL, realized that it must be related to the Price equation. SL and LP explored the applications and implications of the connection. CF identified and developed the connection to ecological orbits and simple harmonic motion via the maternal effect. LP drafted the initial manuscript; both SL and LP edited the first version posted on arXiv \cite{liorsdottir2023virial}. CF, SL, and LP edited the final version and code \cite{pachterlab_FLP2024}. The authors are ordered alphabetically.

\section*{Acknowledgments}

SL thanks Manoj Kaplinghat and Gopolang Mohlabeng who led the ``Introduction to Astrophysics" cluster (cluster 4) at the 2023 UC Irvine COSMOS program. LP relied in part on notes about Fisher's theorem of natural selection from his April 22, 2008 lecture for UC Berkeley course Math 239: Discrete Mathematics for the Life Sciences that were transcribed and edited by Cynthia Vinzant and Caroline Uhler. LP thanks Junhyong Kim for suggesting a possible connection of this work to the ideas in the book \textit{Ecological Orbits: How Planets Move and Populations Grow} by Lev Ginzburg and Mark Colyvan, a discussion that led to CF's extension of the work described in the book and the section on simple harmonic motion.

\bibliographystyle{plain} 
\bibliography{references}

\begin{thebibliography}{10}

\bibitem{alazard2023virial}
Thomas Alazard and Claude Zuily.
\newblock Virial theorems and equipartition of energy for water-waves.
\newblock {\em arXiv preprint arXiv:2304.07872}, 2023.

\bibitem{andersen2004population}
Esben~S Andersen.
\newblock Population thinking, {Price’s} equation and the analysis of
  economic evolution.
\newblock {\em Evolutionary and Institutional Economics Review}, 1:127--148,
  2004.

\bibitem{Bian_2015}
Jiang-Hui Bian, Shou-Yang Du, Yan Wu, Yi-Fan Cao, Xu-Heng Nie, Hui He, and
  Zhi-Bing You.
\newblock Maternal effects and population regulation: maternal density-induced
  reproduction suppression impairs offspring capacity in response to immediate
  environment in root voles microtus oeconomus.
\newblock {\em Journal of Animal Ecology}, 84(2):326--336, 2015.

\bibitem{bijma2020price}
P~Bijma.
\newblock The price equation as a bridge between animal breeding and
  evolutionary biology.
\newblock {\em Philosophical Transactions of the Royal Society B},
  375(1797):20190360, 2020.

\bibitem{bourrat2023price}
Pierrick Bourrat, William Godsoe, Pradeep Pillai, Tarik~C Gouhier, Werner
  Ulrich, Nicholas~J Gotelli, and Matthijs van Veelen.
\newblock What is the price of using the {Price} equation in ecology?
\newblock {\em Oikos}, page e10024, 2023.

\bibitem{clausius1870xvi}
Rudolf Clausius.
\newblock {XVI. On a mechanical theorem applicable to heat}.
\newblock {\em The London, Edinburgh, and Dublin Philosophical Magazine and
  Journal of Science}, 40(265):122--127, 1870.

\bibitem{cresson2021stochastic}
Jacky Cresson, Laurent Nottale, and Thierry Lehner.
\newblock Stochastic modification of newtonian dynamics and induced
  potential—application to spiral galaxies and the dark potential.
\newblock {\em Journal of Mathematical Physics}, 62(7), 2021.

\bibitem{lhospitale1696}
Guillame de~L'H\^{o}pital.
\newblock {\em {Analyse Des Infiniment Petits Pour L'Intelligence Des Lignes
  Courbes}}.
\newblock Chez Montalant, Paris, France, 1696.

\bibitem{einstein1922grundlage}
Albert Einstein.
\newblock {Die Grundlage der allgemeinen Relativit{\"a}tstheorie}.
\newblock {\em Annalen der Physik}, 49(7), 1922.

\bibitem{ellner2011does}
Stephen~P Ellner, Monica~A Geber, and Nelson~G Hairston~Jr.
\newblock Does rapid evolution matter? measuring the rate of contemporary
  evolution and its impacts on ecological dynamics.
\newblock {\em Ecology letters}, 14(6):603--614, 2011.

\bibitem{englert2014lectures}
Berthold-Georg Englert.
\newblock {\em lectures on classical electrodynamics}.
\newblock World Scientific Publishing Company, 2014.

\bibitem{pachterlab_FLP2024}
Catherine Felce, Steinunn Liorsd\'{o}ttir, and Lior Pachter.
\newblock Code to reproduce {Fig.} 1 and {Fig.} 2.:
  \url{https://github.com/pachterlab/FLP_2024}, 2024.
\newblock Accessed: 2024-12-30.

\bibitem{fisher1930genetical}
Ronald~A Fisher.
\newblock {\em The genetical theory of natural selection}.
\newblock Clarendon Press, Oxford, Valorium edition, Bennett JH (Editor), 1999,
  Oxford University Press, Oxford, UK], 1930.

\bibitem{fowler1967statistical}
Ralph~H Fowler.
\newblock {\em {Statistical Mechanics}}.
\newblock Cambridge University Press, Cambridge, UK, 1929.

\bibitem{frank1997price}
Steven~A Frank.
\newblock {The Price equation, Fisher's fundamental theorem, kin selection, and
  causal analysis}.
\newblock {\em Evolution}, 51(6):1712--1729, 1997.

\bibitem{frank1990distribution}
Steven~A Frank and Montgomery Slatkin.
\newblock The distribution of allelic effects under mutation and selection.
\newblock {\em Genetics Research}, 55(2):111--117, 1990.

\bibitem{georgescu1999virial}
Vladimir Georgescu and Christian G{\'e}rard.
\newblock On the virial theorem in quantum mechanics.
\newblock {\em Communications in Mathematical Physics}, 208(2):275--281, 1999.

\bibitem{Ginzburg_Colyvan_2004}
Lev Ginzburg and Mark Colyvan.
\newblock {\em Ecological orbits: How planets move and populations grow}.
\newblock Oxford University Press, 2004.

\bibitem{Ginzburg_94}
Lev~R. Ginzburg and Dale~E. Taneyhill.
\newblock Population cycles of forest {Lepidoptera}: A maternal effect
  hypothesis.
\newblock {\em Journal of Animal Ecology}, 63(1):79--92, 1994.

\bibitem{hanson1995virial}
Mervin~P Hanson.
\newblock The virial theorem, perfect gases, and the second virial coefficient.
\newblock {\em Journal of chemical education}, 72(4):311, 1995.

\bibitem{Harding_2001}
JE~Harding.
\newblock The nutritional basis of the fetal origins of adult disease.
\newblock {\em International Journal of Epidemiology}, 30(1):15--23, 02 2001.

\bibitem{harman2011price}
Oren Harman.
\newblock {\em The price of altruism: {George Price} and the search for the
  origins of kindness}.
\newblock WW Norton \& Company, New York, NY, 2011.

\bibitem{hitchcock2014evolutionary}
Christopher Hitchcock and Joel~D Velasco.
\newblock {Evolutionary and Newtonian forces}.
\newblock {\em Ergo}, 1(2):39, 2014.

\bibitem{liorsdottir2023virial}
Steinunn Liorsd{\'o}ttir and Lior Pachter.
\newblock The virial theorem and the price equation.
\newblock {\em arXiv preprint arXiv:2312.06114 v1}, 2023.

\bibitem{luque2017one}
Victor~J Luque.
\newblock {One equation to rule them all: a philosophical analysis of the Price
  equation}.
\newblock {\em Biology \& Philosophy}, 32(1):97--125, 2017.

\bibitem{luque2021mirror}
Victor~J Luque and Lorenzo Baravalle.
\newblock The mirror of physics: on how the {Price} equation can unify
  evolutionary biology.
\newblock {\em Synthese}, 199(5-6):12439--12462, 2021.

\bibitem{manolio2009finding}
Teri~A Manolio, Francis~S Collins, Nancy~J Cox, David~B Goldstein, Lucia~A
  Hindorff, David~J Hunter, Mark~I McCarthy, Erin~M Ramos, Lon~R Cardon,
  Aravinda Chakravarti, et~al.
\newblock Finding the missing heritability of complex diseases.
\newblock {\em Nature}, 461(7265):747--753, 2009.

\bibitem{marc1985virial}
Guilhem Marc and William~G McMillan.
\newblock The virial theorem.
\newblock {\em Advances in Chemical Physics}, pages 209--361, 1985.

\bibitem{matthen2002two}
Mohan Matthen and Andr{\'e} Ariew.
\newblock Two ways of thinking about fitness and natural selection.
\newblock {\em The Journal of Philosophy}, 99(2):55--83, 2002.

\bibitem{merritt1987distribution}
David Merritt.
\newblock The distribution of dark matter in the {Coma} cluster.
\newblock {\em The Astrophysical Journal}, 313:121--135, 1987.

\bibitem{nourmohammad2013universality}
Armita Nourmohammad, Torsten Held, and Michael L{\"a}ssig.
\newblock Universality and predictability in molecular quantitative genetics.
\newblock {\em Current opinion in genetics \& development}, 23(6):684--693,
  2013.

\bibitem{oguz1990generalization}
Hasan~N Oguz and Andrea Prosperetti.
\newblock A generalization of the impulse and virial theorems with an
  application to bubble oscillations.
\newblock {\em Journal of Fluid Mechanics}, 218:143--162, 1990.

\bibitem{pearson1899vi}
Karl Pearson, Alice Lee, and Leslie Bramley-Moore.
\newblock {VI. Mathematical contributions to the theory of evolution.—VI.
  Genetic (reproductive) selection: Inheritance of fertility in man, and of
  fecundity in thoroughbred racehorses}.
\newblock {\em Philosophical Transactions of the Royal Society of London.
  Series A, Containing Papers of a Mathematical or Physical Character},
  6(192):257--330, 1899.

\bibitem{podio2019virial}
Paolo Podio-Guidugli.
\newblock The virial theorem: A pocket primer.
\newblock {\em Journal of Elasticity}, 137(2):219--235, 2019.

\bibitem{price1972fisher}
George~R Price.
\newblock Fisher's ‘fundamental theorem’ made clear.
\newblock {\em Annals of human genetics}, 36(2):129--140, 1972.

\bibitem{price1970selection}
George~R Price et~al.
\newblock Selection and covariance.
\newblock {\em Nature}, 227:520--521, 1970.

\bibitem{queller2017fundamental}
David~C Queller.
\newblock Fundamental theorems of evolution.
\newblock {\em The American Naturalist}, 189(4):345--353, 2017.

\bibitem{rayleigh1905xlii}
Lord Rayleigh.
\newblock {XLII. On the momentum and pressure of gaseous vibrations, and on the
  connexion with the virial theorem}.
\newblock {\em The London, Edinburgh, and Dublin Philosophical Magazine and
  Journal of Science}, 10(57):364--374, 1905.

\bibitem{reznick1995maternal}
D.~Reznick, H.~Callahan, and R.~Llauredo.
\newblock Maternal effects on offspring quality in poeciliid fishes.
\newblock {\em American Zoologist}, 36(2):147--156, 2015.

\bibitem{rice2004evolutionary}
Sean~H Rice.
\newblock {\em Evolutionary theory: mathematical and conceptual foundations}.
\newblock Sinauer Associates, Sunderland MA, 2004.

\bibitem{rice2008stochastic}
Sean~H Rice.
\newblock A stochastic version of the {Price} equation reveals the interplay of
  deterministic and stochastic processes in evolution.
\newblock {\em BMC evolutionary biology}, 8:1--16, 2008.

\bibitem{Roseboom2006DutchFamine}
Tessa Roseboom, Susanne de~Rooij, and Rebecca Painter.
\newblock The dutch famine and its long-term consequences for adult health.
\newblock {\em Early Human Development}, 82(8):485--491, August 2006.
\newblock Epub 2006 Jul 28.

\bibitem{rudin1964principles}
Walter Rudin et~al.
\newblock {\em Principles of mathematical analysis}, volume~3.
\newblock McGraw-hill New York, 1964.

\bibitem{shore2012introduction}
Steven~N Shore.
\newblock {\em An introduction to astrophysical hydrodynamics}.
\newblock Academic Press, Cambridge, MA, 2012.

\bibitem{simpson1951interpretation}
Edward~H Simpson.
\newblock The interpretation of interaction in contingency tables.
\newblock {\em Journal of the Royal Statistical Society: Series B
  (Methodological)}, 13(2):238--241, 1951.

\bibitem{elliott1984nature}
Elliott Sober.
\newblock {\em The nature of selection}.
\newblock MIT Press, Cambridge, MA, 1984.

\bibitem{suhov2016basic}
Yuri Suhov, Izabella Stuhl, Salimeh Yasaei~Sekeh, and Mark Kelbert.
\newblock Basic inequalities for weighted entropies.
\newblock {\em Aequationes mathematicae}, 90:817--848, 2016.

\bibitem{the1986aj}
Lih~S The and Simon D~M White.
\newblock The mass of the {Coma} cluster.
\newblock {\em Astronomical Journal}, 92(6):1248--1253, 1986.

\bibitem{thorne1976nutrition}
E~Tom Thorne, Ron~E Dean, and William~G Hepworth.
\newblock Nutrition during gestation in relation to successful reproduction in
  elk.
\newblock {\em The Journal of Wildlife Management}, pages 330--335, 1976.

\bibitem{turelli1984heritable}
Michael Turelli.
\newblock Heritable genetic variation via mutation-selection balance: Lerch's
  zeta meets the abdominal bristle.
\newblock {\em Theoretical population biology}, 25(2):138--193, 1984.

\bibitem{van2012group}
Matthijs Van~Veelen, Juli{\'a}n Garc{\'\i}a, Maurice~W Sabelis, and Martijn
  Egas.
\newblock Group selection and inclusive fitness are not equivalent; the {Price}
  equation vs. models and statistics.
\newblock {\em Journal of theoretical biology}, 299:64--80, 2012.

\bibitem{wagner2010measurement}
G{\"u}nter~P Wagner.
\newblock The measurement theory of fitness.
\newblock {\em Evolution}, 64(5):1358--1376, 2010.

\bibitem{walsh2002trials}
Denis~M Walsh, Tim Lewens, and Andr{\'e} Ariew.
\newblock The trials of life: Natural selection and random drift.
\newblock {\em Philosophy of Science}, 69(3):452--473, 2002.

\bibitem{walsh2000chasing}
Dennis~M Walsh.
\newblock Chasing shadows: natural selection and adaptation.
\newblock {\em Studies in History and Philosophy of Science Part C: Studies in
  History and Philosophy of Biological and Biomedical Sciences},
  31(1):135--153, 2000.

\bibitem{williamson2005simultaneous}
Scott~H Williamson, Ryan Hernandez, Adi Fledel-Alon, Lan Zhu, Rasmus Nielsen,
  and Carlos~D Bustamante.
\newblock Simultaneous inference of selection and population growth from
  patterns of variation in the human genome.
\newblock {\em Proceedings of the National Academy of Sciences},
  102(22):7882--7887, 2005.

\bibitem{yule1903notes}
G~Udny Yule.
\newblock Notes on the theory of association of attributes in statistics.
\newblock {\em Biometrika}, 2(2):121--134, 1903.

\bibitem{zhang2010change}
Xu-Sheng Zhang and William~G Hill.
\newblock Change and maintenance of variation in quantitative traits in the
  context of the price equation.
\newblock {\em Theoretical population biology}, 77(1):14--22, 2010.

\bibitem{zwicky1933rotverschiebung}
Fritz Zwicky.
\newblock {Die Rotverschiebung von extragalaktischen Nebeln}.
\newblock {\em Helvetica Physica Acta, Vol. 6, p. 110-127}, 6:110--127, 1933.

\end{thebibliography}
\end{document}